\documentclass[aps,pra,superscriptaddress,twocolumn,longbibliography]{revtex4-1}
\usepackage{bbm}
\usepackage{mathrsfs}
\usepackage{amsmath}
\usepackage{amsfonts}
\usepackage[colorlinks=true,linkcolor=blue,urlcolor=blue,citecolor=blue,anchorcolor=blue]{hyperref}
\usepackage{graphicx,epstopdf}
\usepackage{subfigure}
\usepackage{epsfig}
\usepackage{dcolumn}
\usepackage{bm}
\usepackage{color}
\usepackage{natbib}
\usepackage{amssymb}
\usepackage{xcolor}
\usepackage{braket}
\usepackage{ulem}
\usepackage{float}
\usepackage{braket}
\usepackage{changes}

\begin{document}
	
	\title{Quantum phase transition in a quantum Rabi square with next-nearest-neighbor hopping}
	
	\author{Yilun Xu}
	\affiliation{State Key Laboratory for Mesoscopic Physics, School of Physics, Frontiers Science Center for Nano-optoelectronics, Peking University, Beijing 100871, China}
	\affiliation{Beijing Academy of Quantum Information Sciences, Beijing 100193, China}
	\author{Feng-Xao Sun}
	\email{sunfengxiao@pku.edu.cn}
	\affiliation{State Key Laboratory for Mesoscopic Physics, School of Physics, Frontiers Science Center for Nano-optoelectronics, Peking University, Beijing 100871, China}
	\affiliation{Collaborative Innovation Center of Extreme Optics, Shanxi University, Taiyuan, Shanxi 030006, China}
	\author{Qiongyi He}
	\affiliation{State Key Laboratory for Mesoscopic Physics, School of Physics, Frontiers Science Center for Nano-optoelectronics, Peking University, Beijing 100871, China}
	\affiliation{Collaborative Innovation Center of Extreme Optics, Shanxi University, Taiyuan, Shanxi 030006, China}
	\author{Han Pu}
	\affiliation{Department of Physics and Astronomy, Rice University, Houston, Texas 77251-1892, USA}
	\author{Wei Zhang}
	\affiliation{Department of Physics and Key Laboratory of Quantum State Construction and Manipulation (Ministry of Education), Renmin University of China, Beijing 100872, China}
	\affiliation{Beijing Academy of Quantum Information Sciences, Beijing 100193, China}
	
	\begin{abstract}
		We propose a quantum Rabi square model where both the nearest-neighbor and the next-nearest-neighbor photon hopping are allowed among four quantum Rabi systems located at the vertices of a square. By tuning the next-nearest hopping strength, we realize a first-order phase transition between the antiferromagnetic superradiant phase and the frustrated superradiant phase, as well as a second-order phase transition between the normal and the superradiant phases. To understand the emergence of such phases, we show analytically that the effect induced by next-nearest hopping is equivalent to that of an artificial gauge phase. Our findings suggest that the next-nearest-neighbor hopping can serve as an alternative for the gauge phase to realize quantum control in applications of quantum simulation and quantum materials, and that our model represents a basic building block for the frustrated $J_1$-$J_2$ quantum spin model on square lattices.
	\end{abstract} 
	
	\maketitle
	\section{introduction}
	The study of quantum phase transition (QPT) in systems of light-matter interaction has remained at the frontier of quantum optics and atomic physics for more than a century. 
	A prototypical model supporting QPT is the Dicke model~\cite{HEPP1973360,PhysRevA.7.831,PhysRevLett.90.044101,PhysRevE.67.066203}, where a superradiant phase transition is induced by the strong atom-light interaction and the thermodynamic limit can be satisfied if the atom number $N\rightarrow\infty$.
	Later, QPT has been revealed in the quantum Rabi model (QRM) and the Jaynes-Cummings model~\cite{PhysRevLett.115.180404,PhysRevLett.117.123602}, where a single-mode cavity field and a two-level atom are coupled. 
	With fast development of experimental techniques, significant progresses have been achieved in realizing strong light-matter coupling and controlling related parameters, making it possible to demonstrate QPT in well controlled manners~\cite{PhysRevA.75.013804,PhysRevLett.109.010501,RN102,RN81,RN83,RN84,RN86,PhysRevLett.105.237001,PhysRevLett.102.186402,RN87,RN88,RN69,PhysRevLett.128.160504}. 
	Much interests have been intrigued in the investigation of exotic quantum phases in such models as QRM~\cite{PhysRevLett.127.063602,PhysRevLett.129.183602,PhysRevLett.130.043602,PhysRevLett.115.180404,PhysRevA.101.063843,PhysRevA.108.043705}, Jaynes-Cummings lattice model~\cite{PhysRevLett.117.123602,PhysRevLett.109.053601}, Tavis-Cummings model~\cite{PhysRevLett.124.073602}, Dicke model~\cite{PhysRevLett.120.183603,PhysRevA.85.043821,PhysRevApplied.9.064006,PhysRevLett.128.163601,PhysRevLett.112.173601,PhysRevLett.118.073001,PhysRevResearch.5.L042016}, and so on~\cite{PhysRevB.107.094415,PhysRevA.108.023723,Cheng_2023}. 
	
	Among them, QPTs in coupled few-cavity systems, such as in Rabi dimer (chain)~\cite{PhysRevA.101.063843}, Jaynes-Cummings dimer (chain)~\cite{PhysRevLett.117.123602}, quantum Rabi trimer (ring)~\cite{PhysRevLett.127.063602,PhysRevLett.129.183602} and Dicke trimer~\cite{PhysRevLett.128.163601}, have been proposed to simulate and investigate emergent phenomena in strongly correlated systems. Various intriguing phenomena traditionally explored in condense matter physics can be observed in such light-matter coupled systems. Especially, it is suggested that a one-dimensional coupled array of cavity systems with artificial gauge field can be mapped to a chiral magnetic model consisting of various kinds of magnetic couplings and present rich phase diagrams of magnetic orders~\cite{PhysRevLett.127.063602,PhysRevLett.129.183602}. In such examples, the nontrivial phase of the photon hopping amplitude resulted from the synthetic gauge field, breaks the time-reversal symmetry and plays a crucial role in determining the quantum phases of the system. However, the realization and manipulation of the synthetic gauge field is usually a technically challenging task. Considering the long-lasting interests of investigating quantum magnetic phases, it is then desirable to seek for alternative approaches to realize similar effect without the need of a synthetic gauge field.
	
	
	One promising route to induce exotic phases is through hopping of longer range than nearest neighbors. To this end, the atom-photon platform offers a unique advantage in realizing and manipulating such long-range coupling, which is usually very weak and does not have much room to vary in solids. Long-range coupling has been realized in photonic systems~\cite{madsen_quantum_2022,li_higher-order_2020,Tian:23}, cold atoms~\cite{PhysRevLett.129.220403, PhysRevLett.132.063401}, trapped ions~\cite{RevModPhys.93.025001}, and superconducting qubits~\cite{houck_-chip_2012}. These experimental progresses suggest to use long-range coupling as a useful tool for realizing and studying quantum phases. As we will show here, as a potential substitute for the synthetic gauge field, the quantum Rabi lattices with beyond-nearest-neighbor hopping provides new possibilities to manipulate various quantum phases.
	
\begin{figure}[tb]
		\centering
		\includegraphics[width=0.45\textwidth]{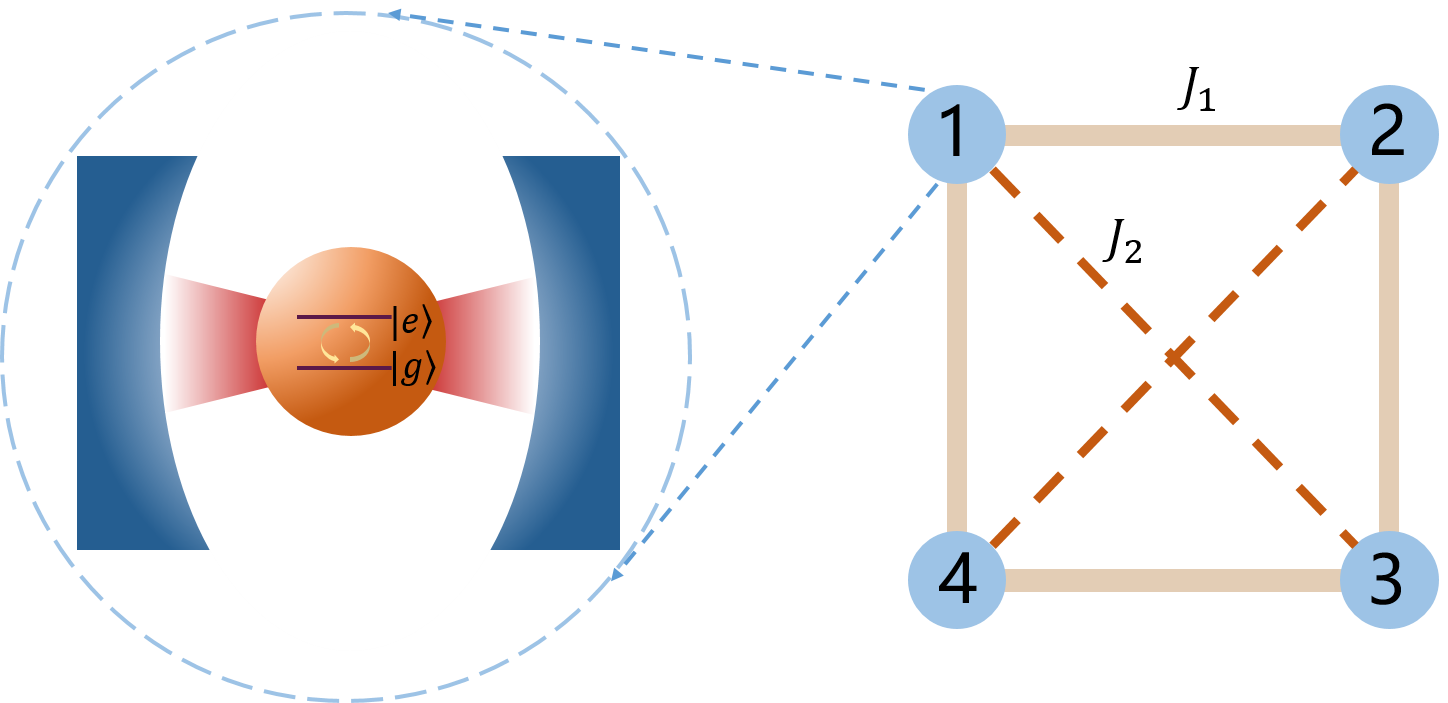}
		\caption{The schematic diagram of quantum Rabi square (QRS) model. Four individual QRMs with identical parameters are located in the vertices of a square lattice, respectively, where the nearest hopping $J_1$ and the next-nearest hopping $J_2$ are included in the lattice.}
		\label{fig_model}
	\end{figure}

	In this paper, we study a quantum Rabi square (QRS) model constructed by four QRMs of interacting two-level atom and cavity photon, residing on the vertices of a square, as schematically depicted in Fig.~\ref{fig_model}. The QRMs are coupled by both the nearest-neighbor and the next-nearest-neighbor photon hopping, with respective hopping amplitudes $J_1$ and $J_2$. We consider both hopping parameters to be of real values and hence there is no artificial gauge field present. From a different perspective, our model can also be regarded as a building block for the $J_1$-$J_2$ spin Heisenberg model~\cite{PhysRevB.41.9323,PhysRevB.44.12050,PhysRevB.54.9007} which is extensively studied in quantum magnetism, hence can shed light on the implementation and investigation of interacting spin models.
 
 By treating the cavity photon via a mean-field approximation, we obtain analytic solution of the ground state, and map out the phase diagram by tuning the atom-light coupling and the relative hopping strength of $J_2/J_1$. When the atom-light coupling is weak, all QRMs are in the normal phase (NP) with zero expectation of cavity photon. By increasing the coupling strength across the critical point, the system will go through a second-order phase transition to enter a superradiant phase (SRP) along with the spontaneous breaking of the $Z_2$ symmetry of photons and geometric $C_4$ symmetry of the square. The SRP can be further divided into two branches distinguished by the symmetry of the ground state when tuning the next-nearest-neighbor hopping. The two different branches of SRP can be mapped to (anti)ferromagnetic and frustrated phases of quantum magnetic models, which are also observed in the quantum Rabi ring (QRR) model involving artificial gauge phase~\cite{PhysRevLett.127.063602, PhysRevLett.129.183602}. Finally, we derive an analytic correspondence between the next-nearest-neighbor hopping and gauge field for this model. Our work suggests the usage of long-range coupling as a versatile tool in the simulation and study of QPTs.

	\section{Quantum Rabi Square model}
	
	We consider a QRS model constructed by four identical QRMs sitting on the vertices of a square, as shown in Fig.~\ref{fig_model}. Each QRM is coupled with its two nearest neighbors with a photon hopping amplitude $J_1$ (edges of the square), and with the opposite node with hopping amplitude $J_2$ (diagonals). The system Hamiltonian reads
	\begin{align}
		\label{original Hamiltonian}
		H&=\sum_{i=1}^{4}H_{R,i}+H_{\rm hop}+H_{\rm next}, \nonumber \\
		H_{R,i}&=\omega a_i^\dagger a_i+\frac{\Omega}{2}\sigma_i^z+\lambda(a_i^\dagger+a_i)\sigma_i^x, \nonumber \\
		H_{\rm hop}&=\sum_{i=1}^4J_1a_ia_{i+1}^\dagger+ {\rm H.c.}, \nonumber \\
		H_{\rm next}&=J_2(a_1a_3^\dagger+a_2a_4^\dagger+ {\rm H.c.}).
	\end{align}
	In the Hamiltonian of the QRM on the $i$-th site $H_{R,i}$, $a_i$ and $a_i^\dagger$ are the field operators for the optical field, and $\sigma_i^z$ stands for the Pauli spin operator in the $z$-direction for the two-level atom. The four QRMs have the same photon frequency $\omega$, the energy gap of two-level atom $\Omega$, and the cavity-atom coupling strength $\lambda$. The general scenario of non-identical nodes can be investigated analogously. The nearest-neighbor and next-nearest-neighbor hopping of photons are represented by $H_{\rm hop}$ and $H_{\rm next}$, respectively. Notice that here we consider only real values for both $J_1$ and $J_2$. Thus, the system does not possess any gauge field which is the main focus of the previously studied quantum Rabi ring model~\cite{PhysRevLett.127.063602,PhysRevLett.129.183602}.
	
	When all the couplings between the four sites are zero, the system consists of four identical and isolated QRMs, which can host a QPT between NP and SRP~\cite{PhysRevLett.115.180404}, in the so-called classical oscillator limit, i.e., $\Omega/\omega\rightarrow\infty$. This is because the macroscopic photonic occupation of the cavity mode is proportional to $\Omega/\omega$. By turning on a finite nearest-neighbor hopping $J_1$, the system can present a richer phase diagram with multiple superradiant phases showing magnetic and chiral signatures~\cite{PhysRevLett.129.183602}. A recent theoretical study has involved the next-nearest-neighbor coupling $J_2$ for a one-dimensional Dicke lattice model~\cite{PhysRevResearch.5.L042016}. However, the effect of such long-range hopping on the SRPs is still an open question.
	
	
	Before showing our main results in detail, we will analyze the symmetry of the QRS model briefly. The parity symmetry of the total excitation number is preserved in the QRS, which is isomorphic to the $Z_2$ symmetry. Defining the parity operator as $\hat{P}=\exp(i\pi\sum_{i=1}^{4}\hat{N}_i)$, where $\hat{N}_i\equiv a_i^\dagger a_i+\sigma_i^+\sigma_i^-$ represents the excitation number of the $i$-th cavity, the commutation relation $[H,\hat{P}]=0$ can be easily verified. The two eigenvalues of the parity operator are $\pm1$, denoting the vectors in subspace with even (odd) total excitations. Furthermore, the cyclic symmetry is also present in this Hamiltonian. This means that the system remains unchanged by rotating the four sites as $1234\rightarrow2341\rightarrow3412\rightarrow4123$, which composes the element of the $C_4$ group. Thus, the Hamiltonian $H$ follows a total $Z_2\times C_4$ symmetry.
	
	\section{Normal phase}\label{section}
	
	With the aid of the Schrieffer-Wolff transformation, the original Hamiltonian (\ref{original Hamiltonian}) can be simplified. The idea of the Schrieffer-Wolff transformation is to rotate the system into another representation by a well designed unitary operator $U=e^S$, where $S$ is an anti-Hermitian operator and the coupling between the spin up and spin down subspaces can be effectively eliminated by the term $[H,S]$. The transformation matrix can be expressed as $U=\Pi_{n}U_n$, where $U_n=\exp[-i\frac{g}{\sqrt{\eta}}\sigma_n^y(a_n^\dagger+a_n)]$. Here, the two key dimensionless parameters are defined as \[g\equiv{\lambda}/{\sqrt{\Omega\omega}}\,,\;\;\;\eta\equiv\Omega/\omega\,,\] representing the normalized atom-cavity coupling strength and the ratio of the atomic and the photon frequencies, respectively. The normalized coupling strength is kept finite in our discussion, and the hopping strengths among the sites are much smaller than the frequencies of the optical modes. In summary, the parameters of the model are assumed to satisfy $\Omega\gg\lambda\gg\omega\gg J_1,J_2$. 
	
	Taking this condition into account, the effective Hamiltonian after the Schrieffer-Wolff transformation can be calculated through $H^{\rm eff}=e^{-S}He^S$, resulting in
	\begin{align}
		H^{\rm eff}&= \sum_{i=1}^{4}H_{R,i}^{\rm eff}+H_{\rm hop}+H_{\rm next}+\mathcal{O}(\omega \eta^{-1}), \nonumber \\
		H_{R,i}^{\rm eff}&=\omega a_i^\dagger a_i +\frac{\Omega}{2}\sigma_i^z+\omega g^2\sigma_i^z(a_i+a_i^\dagger)^2.
	\end{align}
	After projecting the effective Hamiltonian into the four-fold spin down subspace, {\it i.e.}, $H^{\rm eff}_\downarrow\equiv \text{tr}[(\ket{\downarrow}\bra{\downarrow})^{\otimes4}H^{\rm eff}]$, the Hamiltonian will take the form of
	\begin{align}\label{Hamiltonian_SW}
		H^{\rm eff}_\downarrow&= \sum_{i=1}^{4}(\omega-2\omega g^2)a_i^\dagger a_i-\omega g^2(a_i^2+a_i^{\dagger2}) \nonumber\\
		&\hspace{1em}+\sum_{i,j=1}^{4}a_i^\dagger M_{ij}a_j+E_0, \nonumber\\
		E_0&=4 \left(-\dfrac{\Omega}{2}-\omega g^2+\dfrac{\omega^2 g^2}{\Omega} \right), \nonumber\\
		M&=\begin{pmatrix}
			0&J_1&J_2&J_1\\
			J_1&0&J_1&J_2\\
			J_2&J_1&0&J_1\\
			J_1&J_2&J_1&0
		\end{pmatrix}.
	\end{align}
	The term $\sum_{i,j=1}^{4}a_i^\dagger M_{ij}a_j$ originates from the nearest and next-nearest neighbor hopping interaction $H_{\rm hop}+H_{\rm next}$. 
 
 Since the translation invariance is preserved in this system, the Bloch theorem can be applied. To this end, we transform the creation and annihilation operators into momentum space as $a_n^\dagger=({1}/{\sqrt{N}})\sum_{n=1}^Ne^{inq}a_q^\dagger$, where $q={2\pi l}/{N}$ is the system momentum, and $l = 0, 1, ..., N-1$ with $N=4$ for our QRS model. By means of this transformation, the Hamiltonian in momentum space can be expressed as
	\begin{align}
		H^{\rm eff}_\downarrow&=\sum_{q} \left[\omega_qa_q^\dagger a_q-\omega g^2(a_qa_{-q}+a_q^\dagger a_{-q}^\dagger) \right]+E_0,
	\end{align}
	where $\omega_q=\omega-2\omega g^2+J_2\cos(2q)+2J_1\cos(q)$. Then, using the two-mode squeezing transformation, {\it i.e.}, the Bogoliubov transformation $S_q=\exp[\lambda_q(a_q^\dagger a_{-q}^\dagger-a_qa_{-q})]$ with $\lambda_q=\dfrac{1}{8}\ln\dfrac{\omega_q+\omega_{-q}+4\omega g^2}{\omega_q+\omega_{-q}-4\omega g^2}$, we can obtain the ground state energy $E_g$ and the excitation energy $\varepsilon_q$,
	\begin{align}
		E_g&=E_0+\dfrac{1}{2}\sum_{q}(\varepsilon_q-\omega_q), \nonumber \\
		\varepsilon_q&=\dfrac{1}{2} \left[\sqrt{(\omega_q+\omega_{-q})^2-16\omega^2g^4}+\omega_q-\omega_{-q}\right],
		\label{eq:excited-energy}
	\end{align}
	and write the effective Hamiltonian as $H^{\rm eff}_\downarrow=\sum_{q}\varepsilon_qa_q^\dagger a_q+E_g$.
	The critical points will be reached when the excited energy $\varepsilon_q$ vanishes, giving
	\begin{align}
		\label{QRM_critical_point}
		4g_c^2(q)=1+\frac{J_2}{\omega}\cos(2q)+\frac{2J_1}{\omega}\cos(q).
	\end{align}
	It is straightforward to verify that $g_c({\pi}/{2})=g_c({3\pi}/{2})$ and $g_c(\pi)<g_c(0)$ when $J_1>0$. Thus, we only need to focus on the two branches of the SRP denoted by $q=\pi$ and $q=\pi/2$ $(3\pi/2)$ for $J_1>0$. For the case of $J_1<0$, a similar analysis can be carried out and will be briefly discussed later.
	
	\section{Superradiant phase}
	
	When the dimensionless coupling strength $g$ exceeds the critical point $g_c$, the first-excited energy derived in the previous section [{i.e.}, Eq.~(\ref{eq:excited-energy})] will be revised, and the superradiant phase transition occurs. By treating the optical mode via a mean-field approach, the annihilation operator can be rewritten as $a_n\rightarrow a_n+\alpha_n$, where the mean value of the optical field amplitude is complex and denoted as $\alpha_n=A_n+iB_n$. Then the Hamiltonian becomes
	\begin{align}
		\label{QRM_SRP}
		H&=\sum_n \Big[\omega a_n^\dagger a_n+\dfrac{\Omega_n}{2}\tau_n^z+\lambda_n(a_n+a_n^\dagger)\tau_n^x\notag\\
		& \hspace{0.5cm} +J_1a_n^\dagger(a_{n+1}+a_{n-1})+J_2a_n^\dagger a_{n+2} \Big] + V+E_0,
		\nonumber \\
		E_0&=\sum_n\left|\alpha_n\right|^2+J_1\sum_n\alpha_n^*(\alpha_{n+1}+\alpha_{n-1})\notag\\
		& \hspace{0.5cm} +J_2\sum_n\alpha_n^*\alpha_{n+2},
		\nonumber \\
		V&=\sum_n \Big\{\omega(\alpha_na_n^\dagger+\alpha_n^*a_n)+\lambda \sin(2\gamma_n)\tau_n^z(a_n^\dagger+a_n)\notag\\
		& \hspace{0.5cm} +J_1 \left[a_n^\dagger(\alpha_{n+1}+\alpha_{n-1})+{\rm H.c.} \right]
		\notag\\
		& \hspace{0.5cm} +J_2(a_n^\dagger\alpha_{n+2}+{\rm H.c.}) \Big\},
	\end{align} 
	where the transformed Pauli-Z operator $\tau_n^z=(\Omega\sigma_n^z+4A_n\lambda\sigma_n^x)/\Omega_n$, and $\Omega_n=\sqrt{\Omega^2+16\lambda^2A_n^2}$. Then the eigenvectors of $\tau_n^z$ can be obtained as
	\begin{align}
		\ket{+}&=\cos(\gamma_n)\ket{\uparrow}+\sin(\gamma_n)\ket{\downarrow}, \nonumber\\
		\ket{-}&=-\sin(\gamma_n)\ket{\uparrow}+\cos(\gamma_n)\ket{\downarrow},
	\end{align}
	where $\tan(2\gamma_n)=4\lambda A_n/\Omega$.
	
	Typically, the local minimum can be determined by demanding  $V=0$, which will give us the concrete value for the optical displacement $\left\{\alpha_n\right\}$ in Appendix~\ref{appendix A}. Then, the Hamiltonian retains only two parts in the following form 
	\begin{align}
		H&=\sum_n \Big[\omega a_n^\dagger a_n+\dfrac{\Omega_n}{2}\tau_n^z+\lambda_n(a_n+a_n^\dagger)\tau_n^x\notag\\
		& \hspace{0.5cm} +J_1a_n^\dagger(a_{n+1}+a_{n-1})+J_2a_n^\dagger a_{n+2} \Big]+E_0.
		\label{superradiant Hamiltonian}
	\end{align}
	Notice that the mean-field Hamiltonian Eq.~(\ref{superradiant Hamiltonian}) acquires a similar form as the original Hamiltonian~(\ref{original Hamiltonian}). Thus, we can employ the same approach to obtain the ground state energy, which reads
	\begin{align}
		E_g=&\sum_n \Big\{\omega(A_n^2+B_n^2)+2J_1[(A_nA_{n+1}+B_nB_{n+1})]\notag\\
		&+J_2(A_nA_{n+2}+B_nB_{n+2})-\dfrac{1}{2}\Omega_n \Big\}\notag\\
		&+\sum_n \left(-\omega g_n^2+\dfrac{\omega^2g_n^2}{\Omega_n}\right)+\dfrac{1}{2}\sum_q(\varepsilon_q'-\omega_q').
	\end{align}
	Here, $g_n={\lambda_n}/{\sqrt{\omega\Omega_n}}$ and $\lambda_n={\lambda\Omega}/{\Omega_n}$. By minimizing the first two lines of $E_g$ which serves as the main contribution to the ground state energy, we can also get the same solutions for $A_n$ and $B_n$ as the ones in Appendix~\ref{appendix A}. In the QRS, $g_n$ and $\Omega_n$ are independent on the site index $n$, so we denote all the $g_n$ and $\Omega_n$ as $g'$ and $\Omega'$ in the following. The ground state energy for all three branches is summarized as
	\begin{align}
		\label{QRM_SRP_ground_energy}
		E_g&=-\left[\dfrac{\lambda^2}{g_c^2(q_0)\omega}+\dfrac{\Omega^2}{\lambda^2}g_c^2(q_0)\omega\right]\notag\\
		&\hspace{0.5cm}+4\left(-\omega g'^2+\dfrac{\omega^2g'^2}{\Omega'}\right)+\dfrac{1}{2}\sum_q(\varepsilon_q'-\omega_q').
	\end{align}
	Here, $\varepsilon_q'=\dfrac{1}{2}[\sqrt{(\omega_q'+\omega_{-q}')^2-16\omega^2g'^4}+\omega_q'-\omega_{-q}']$ with $\omega_q'=\omega-2\omega g'^2+J_2\cos(2q)+2J_1\cos(q)$ being the excitation energy of SRP, which acquires a similar form as in NP with $g$ replaced by $g'$. Note that $g'$ can be expressed in terms of $g$ and $g_c$ as $g'={g_c^3(q_0)}/{g^2}$, if we combine the definition $g'={\lambda'}/{\sqrt{\omega\Omega'}}$ with the following identities
	\begin{align}
		\Omega'&=\sqrt{\Omega^2+16\lambda^2A^2},\nonumber \\
		A^2&=\dfrac{1}{16\lambda^2}\left(\dfrac{16\lambda^4}{[4\omega g_c^2(q_0)]^2}-\Omega^2\right),
		\nonumber \\
		\lambda'&=\lambda_n=\dfrac{\lambda\Omega}{\Omega'}.
	\end{align}

\begin{figure}[tb]
		\centering
		\includegraphics[width=0.5\textwidth]{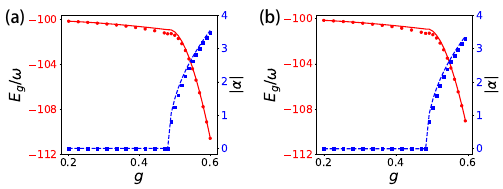}
		\caption{The ground state energy $E_g$ (red solid line with dots) and the order parameter $|\alpha|$ (blue dashed line with squares) are shown in subfigure with (a) $J_2=0.02$; (b) $J_2=0.07$. The numerical results (dots and squares) agree well with the analytical results (red solid and blue dashed lines). We set the fixed parameters $\Omega=50$, $J_1=0.05$, and $\omega=1$ as the unit of energy in all the figures of this paper.}
		\label{order parameter}
	\end{figure}

	The ground state energy $E_g$ and the order parameter $\left|\alpha\right|$ are depicted in Fig.~\ref{order parameter}, where $J_2$ are set at 0.02 and 0.07 in Figs.~\ref{order parameter}(a) and ~\ref{order parameter}(b), respectively. In both cases, numerical simulations (red dots and blue squares) are obtained for comparison by diagonalizing the displaced Hamiltonian $D(\vec{\alpha})HD^\dagger(\vec{\alpha})$, where $H$ is the original Hamiltonian~(\ref{original Hamiltonian}), $\vec{\alpha}\equiv[\alpha_1,\alpha_2,\alpha_3,\alpha_4]^T$, and $D(\vec{\alpha})\equiv\Pi_{n}D(\alpha_n)=\Pi_{n}\exp(\alpha_n^*a_n-\alpha_na_n^\dagger)$ stands for the total displacement operator. The truncated dimension of all four bosonic modes is $N_c=5$ to guarantee convergence. These numerical results agree well with the analytical solutions (red solid and blue dashed lines). Additionally, the first-order derivative of the order parameter and the second-order derivative of the ground state energy are all discontinuous around the critical points, indicating a second-order QPT. The accuracy of the analytical mean-field ground state is analyzed in Appendix~\ref{appendix} by comparing with numerical results obtained by exact diagonalization.

\begin{figure}[tb]
		\centering
		\includegraphics[width=0.5\textwidth]{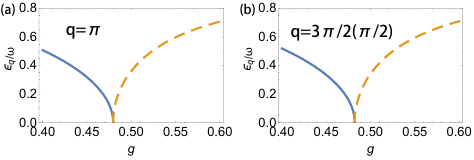}
		\caption{The excited energy scaling around the critical point with (a) $J_2=0.02$ and (b) $J_2=0.07$. Here, the solid lines correspond to the normal phase, and the dashed lines represent the superradiant phase, with the branch of (a) $q=\pi$ and (b) $q=\pi/2$ ($3\pi/2$) dominating.}
		\label{QRM_scaling}
	\end{figure}

	To characterize the QPT between NP and SRP, next we analyze the scaling behavior of the excitation energy near the critical point $g=g_c(q_0)$. It is easy to verify that $\omega_q=\omega_{-q}$ for NP, and $\omega_q'=\omega_{-q}'$ for all possible SRP branches with $q=\{0, \pi/2, \pi, 3\pi/2\}$. Thus, the excited energy in NP and SRP can be reduced to $\varepsilon_q=\sqrt{\omega_q^2-4\omega^2g^4}$ and $\varepsilon_q'=\sqrt{\omega_q'^2-4\omega^2g'^4}$, respectively. Taking NP as an example, we can expand the two terms in the square root up to linear order as $\omega_q^2(g)\sim\omega_q^2(g_c)+k_1(g-g_c)$ and $4\omega^2g^4\sim 4\omega^2g_c^4+k_2(g-g_c)$, where $k_1<0$ and $k_2>0$. The combination of them thus gives a nonzero linear coefficient $(k_1-k_2)$ in the expansion of $\varepsilon_q$. The same analysis can also be applied to SRP and lead to a linear term as well. With that,  the excited energy around $g\rightarrow g_c$ scales as $\varepsilon_q,\varepsilon_q’\propto\left|g-g_c\right|^{1/2}$. The behaviors of excited energy around the critical point are demonstrated in Fig.~\ref{QRM_scaling}, where the SRP branch in Fig.~\ref{QRM_scaling}(a) is $q=\pi$ with $J_2=0.02$ , and  $q=\pi/2$ $(3\pi/2)$ with $J_2=0.07$ in Fig.~\ref{QRM_scaling}(b). In both cases, a typical $1/2$ scaling rule is clearly observed.

	\section{phase diagram and connection to quantum magnetism}
	
	To analyze the phase diagram of the different SRP branches, we follow the discussion of Ref.~\cite{PhysRevLett.129.183602} to consider the real part $A_n$ and imaginary part $B_n$ of the mean-field amplitude of the optical mode as two different directions of the spin operator. As shown in Appendix~\ref{appendix B}, the effective low energy Hamiltonian~(\ref{Hamiltonian_SW}) can be mapped into a spin $J_1$-$J_2$ model. With that, the SRP branches with different choices of $q$ can be characterized by the structure of four ``local spins", and linked to different phases of quantum magnetism. The optical displacements also match well with the spin arrangements.  Specifically, in the $q=0$ branch of SRP, the spins are mostly aligned, being analogous to the ferromagnetic superradiant phase (FRP). In the $q=\pi$ branch, neighboring spins are anti-aligned to demonstrate features of the antiferromagnetic superradiant phase (AFRP). The $q=\pi/2$ $(3\pi/2)$ branch presents spin structure with competing aligned and anti-aligned tendencies, thus can be referred as a frustrated magnetic phase.
	
\begin{figure}[tb]
		\centering
		\includegraphics[width=0.48\textwidth]{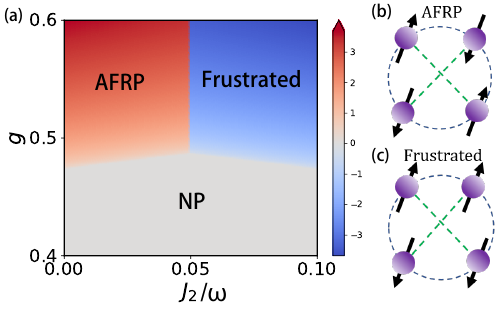}	
		\caption{(a) The phase diagram with a fixed value of the nearest coupling strength $J_1=0.05$. The bottom region (grey) corresponds to the NP, the upper left region (red) to the AFRP, and the upper right region (blue) to the frustrated SRP. The spin directions are displayed when the system is located in the (b) AFRP and (c) frustrated phase, respectively.}
		\label{QRM_phase_diagram_J2}
	\end{figure}

	
	In Fig.~\ref{QRM_phase_diagram_J2}(a), we depict the phase diagram against the next-nearest-neighbor hopping $J_2$ and the atom-cavity coupling $g$, for a fixed nearest-neighbor hopping $J_1=0.05$. Three phases can be clearly found in this phase digram, with the background color denoting $\alpha_n\alpha_{n+2}/\left|\alpha\right|$. Thus the positive and negative values represent the AFRP and frustrated SRP phases, respectively. For small $g$, the system is in the NP region as expected. As $g$ goes beyond the critical value, the system crosses a second-order QPT to become SRP. When $J_2<J_1$, the $q=\pi$ branch is stabilized and the spins are organized antiferromagnetically to form an antiferromagnetic superradiant phase (AFRP), as shown in Fig.~\ref{QRM_phase_diagram_J2}(b). When $J_2 > J_1$, the $q=\pi/2\, (3\pi/2)$ branch is favored to present frustrated spin structures as illustrated in Fig.~\ref{QRM_phase_diagram_J2}(c). The physical mechanism of the spin frustration can be understood as follows. When the next-nearest-neighbor coupling $J_{2}$ is large enough, the spins connected by diagonal lines tend to be anti-aligned. Meanwhile, the nearest-neighbor hopping also favors anti-aligned spins connected by edges. These constraints clearly lead to a frustrated spin configuration with both $Z_2$ and $C_4$ symmetries broken. As a four-site lattice, the emergency of this ``frustrated" phase also serves as a strong evidence for the true frustration effect as the number of sites increases towards infinity. The two SRP branches are separated by a first-order QPT with abrupt change of spin arrangement. If the nearest-neighbor hopping $J_1$ is negative, the phase diagram has the form as  Fig.~\ref{QRM_phase_diagram_J2}(a), with the AFRP region being replaced by ferromagnetic superradiant phase (FRP) since the $q=0$ branch is more favorable.

	\section{the connection to QRR model with artificial gauge field}
	\label{sec:equivalent}
	
	Next, we discuss the connection between the QRS model with next-nearest-neighbor coupling and the QRR model with artificial gauge field. The the latter is obtained from our model by taking $J_2=0$ but making $J_1 \rightarrow J_1^0 \,e^{i\theta}$ complex. It has been shown that the QRR with artifical gauge field can be mapped to a magnetic system of coupled Lipkin-Meshkov-Glick model~\cite{PhysRevLett.129.183602}, and has potential application in the simulation and study of magnetic orders. In the following, we show analytically that by using the next-nearest-neighbor coupling instead of gauge phase, both the energy spectrum and eigenfunctions can be exactly recovered.

	
	
	\subsection{Excitation energy}
	
	We first focus on the excitation energy and derive a mapping between the QRS and QRR models. To distinguish these two models, a superscript $0$ is used to denote the physical quantities already been introduced for the QRS model in the QRR model, without further specification. The excitation energies for NP ($\varepsilon_q^0$) and SRP ($\varepsilon_q'^0$) in the QRR model are given as
	\begin{align}
		\varepsilon_q^0&=\dfrac{1}{2}[\sqrt{(\omega_q^0+\omega_{-q}^0)^2-16\omega^2g^4}+\omega_q^0-\omega_{-q}^0],
		\nonumber \\
		\varepsilon_q'^0&=\dfrac{1}{2}[\sqrt{(\omega_q'^0+\omega_{-q}'^0)^2-16\omega^2(g'^0)^4}+\omega_q'^0-\omega_{-q}'^0],
		\label{excited energy SRP QRR}
	\end{align}
	where $\omega_q^0=\omega-2\omega g^2+2J_1^0\cos(q-\theta)$, $\omega_q'^0=\omega-2\omega (g'^0)^2+2J_1^0\cos(q-\theta)$, and the renormalized coupling strength can be expressed by the critical points in the QRR model as $g'^0=(g_c^0)^3(q_0)/g^2$. Thus, the critical points for different $q$ read
	\begin{align}
		\label{QRM_critical_point_0}
		&4(g_c^0)^2(q)\nonumber\\
		&=\dfrac{1+\frac{4J_1^0}{\omega}\cos(q)\cos(\theta)+\frac{4(J_1^0)^2}{\omega^2}\cos(q-\theta)\cos(q+\theta)}{1+\frac{2J_1^0}{\omega}\cos(q)\cos(\theta)}.
	\end{align}
	The relation between the QRS and QRR models can be established by demanding the excitation energies therein as equal. Because the critical points are also derived by setting the excitation energy to zero, the critical points and resulting phase diagram can be naturally reproduced. 
	
	(i) \textit{The correspondence for AFRP.} By setting $q=\pi$, we reach the following conditions in NP and SRP, respectively,  
	\begin{align}
		\varepsilon_\pi^0(\theta)&=\varepsilon_\pi(J_2),
		\nonumber \\
		\varepsilon_\pi'^0(\theta)&=\varepsilon_\pi'(J_2).
	\end{align}
	Therefore, we can get the constraints for $\theta$ and $J_2$, leading to
	\begin{align}
		\label{corresponding pi}
		J_2=2[J_1-J_1^0\cos(\theta)].
	\end{align}
	This relation establishes an exact connection between the next-nearest-neighbor hopping $J_2$ and gauge phase $\theta$.
	
	(ii) \textit{The correspondence for frustrated SRP.} Since the cases of $q=3\pi/2$ and $\pi/2$ are equivalent, in the following we set $q = 3\pi/2$ for simplicity. The matching conditions for NP and SRP are 
	\begin{align}
		\varepsilon_{3\pi/2}^0(\theta)&=\varepsilon_{3\pi/2}(J_2),
		\nonumber \\
		\varepsilon_{3\pi/2}'^0(\theta)&=\varepsilon_{3\pi/2}'(J_2).
	\end{align} 
	And we can obtain the condition of parameters for the two phases as
	\begin{align}
		J_2&=1-2g^2-\sqrt{(\sqrt{1-4g^2}-2J_1^0\sin(\theta))^2+4g^4},
		\nonumber \\
		\label{corresponding2 3pi/2}
		J_2&=1-2g'^2-\sqrt{(\sqrt{1-4g'^2}-2J_1^0\sin(\theta))^2+4g'^4}.
	\end{align}
	Notice that here the connection between $J_2$ and $\theta$ are of different form for NP and SRP, which join at the critical point to form a continuous function of the dimensionless atom-cavity coupling strength $g$. The value of $J_2$ at the critical point is obtained as $J_2(g=g_c)=4(J_1^0)^2\sin^2(\theta)$, by inserting $g=g_c(3\pi/2)=\frac{1}{2}\sqrt{1-4(J_1^0)^2\sin^2(\theta)}$. 
	
	(iii) \textit{The boundary for the first-order phase transition.} In QRR model with $J_2=0$, the boundary between the $q=\pi$ and $q=\pi/2$ $(3\pi/2)$ phases is given by the intersection of the corresponding critical curves, and the coordinate for the intersection $\theta_c$ can be determined by the following equation
	\begin{align}
		\label{theta_c}
		1-2J_1^0\cos(\theta_c)=1-4(J_1^0)^2\sin^2(\theta_c).
	\end{align}
	In QRS model, the critical point for the first-order phase transition is given by $J_{2c}=J_1$. Combining Eqs.~(\ref{corresponding pi}) and (\ref{corresponding2 3pi/2}), it can be derived that
	\begin{align}
		J_1&=J_{2c}=2[J_1-J_1^0\cos(\theta_c)],
		\nonumber \\
		J_1&=J_{2c}(g=g_c)=4(J_1^0)^2\sin^2(\theta_c).
	\end{align}
	The value of the nearest hopping rate $J_1$ hence can be determined as $J_1=2J_1^0\cos(\theta_c)$ by the first condition, and $J_1=4(J_1^0)^2\sin^2(\theta_c)$ by the second condition. The two solutions are identical according to Eq.~(\ref{theta_c}).
	
	(iv) \textit{The case of negative $J_1$ and FRP.} If we set the nearest hopping strength $J_1<0$, the FRP region can also be simulated in our model with a similar analysis. In such a case, the AFRP vanishes, while the first-order phase transition between the FRP and the frustrated SRP can also be observed. And all the discussions above are still valid after substituting $J_1$ and $\theta$ by $-J_1$ and $\pi-\theta$, respectively.
	
	\subsection{Order parameter}
	
	We can also establish a mapping between QRS and QRR models by requiring the same order parameter $A^2$, i.e., displacement for optical modes. In both models, $A^2$ writes
	\begin{align}
		A^2&=\dfrac{1}{16\lambda^2}\left(\dfrac{16\lambda^4}{[4\omega g_c^2(q_0)]^2}-\Omega^2\right).
	\end{align}
	Obviously, the same critical point is obtained in this mapping. According to Eq.~(\ref{QRM_critical_point_0}), the condition for AFRP and frustrated SRP are achieved as
	\begin{align}
		{\rm AFRP}:~~J_2&=2[J_1-J_1^0 \cos(\theta)],
		\nonumber \\
	\label{order_para_cor_Frustrated}	
	{\rm frustrated}:~~J_2&=4(J_1^0)^2\sin^2(\theta),
	\end{align}
	respectively. In order to acquire the same triple point, we also require $J_{2c}=2[J_1-J_1^0 \cos(\theta_c)]=J_1$, giving $J_1=2J_1^0 \cos(\theta_c)$, or equivalently, $J_{2c}=4(J_1^0)^2\sin^2(\theta_c)=J_1$. Similar mapping for SRP can be reached by setting the negative nearest hopping rate $J_1<0$. We emphasize that while the relation between $J_2$ and $\theta$ in AFRP is the same as the one derived from excitation energy, for frustrated SRP Eq.~(\ref{order_para_cor_Frustrated}) is different from that in Eq.~(\ref{corresponding2 3pi/2}).

	To summarize, the previously studied QRR model with nearest-neighbor hopping strength $J_1^0$ and the gauge phase $\theta$ can be equivalently investigated by using our QRS model without any gauge phases. In order to obtain the same excitation energy, $J_2$ is determined by Eqs.~(\ref{corresponding pi}) and (\ref{corresponding2 3pi/2}). When it comes to the order parameter, the matching conditions give Eq.~(\ref{order_para_cor_Frustrated}). These two considerations give the same requirement of $J_1=2J_1^0\cos(\theta_c)$ to reach the same triple point and phase diagram. Here, $\theta_c$ is obtained through $2J_1^0\cos(\theta_c)=4(J_1^0)^2\sin^2(\theta_c)$. From this aspect, the next-nearest-neighbor hopping can play the same role as the artificial gauge phase. Considering the fact that an experimental realization of synthetic gauge field usually requires a fine-tune of hopping parameter via some complex schemes, e.g., by applying Floquet engineering~\cite{PhysRevA.84.043804,PhysRevLett.127.063602}, and may introduces side effects such as heating and noises, our work provides another route to implement and explore exotic phases in certain experimentally friendly platforms~\cite{PhysRevA.86.023837,PhysRevLett.128.160504}, which involve only real hopping amplitudes.
	
	\section{summary}
	In this paper, we propose a quantum Rabi square model where the effect of the next-nearest neighbor hopping strength is mainly investigated.  We obtain the analytical ground state energy and the critical points for both the first- and the second-order phase transitions. The equivalent spin arrangement of the optical modes will go through a sudden change when the next-nearest hopping strength exceeds the nearest hopping rate. Comparing with the previous work on the QRR model~\cite{PhysRevLett.129.183602}, we find our model provides an alternative approach to realize the global gauge phase by means of the next-nearest hopping strength. The one-to-one corresponding relations between $\theta$ in QRR and $J_2$ in QRS are also obtained from the excited energy and order parameter, respectively. With these results, it can be concluded that the gauge phase plays the same role as an appropriate additional system parameter, such as the next-nearest hopping rate $J_2$ in our QRS model, which raises a crucial explanation on how the gauge phase induces exotic quantum phases. Those extra degrees of freedom will lead to staggered critical curves of the SRP branches. Among them only one of the SRP branches can be revealed without these extra controllable parameters. And the intersections of these critical curves correspond to the triple points of the system, where both the first- and the second-order phase transitions occur. Such a comparison between the QRR model and the QRS model in our work will pave an avenue for seeking exotic quantum phase transitions in matter-light coupled systems, and provide an alternative method to investigate emergent phenomena induced by gauge phases in quantum materials. 

	Finally, we want to comment that our QRS model can be regarded as a basic building block for a lattice system with periodic boundary condition. The connection and difference between artificial gauge phase and long-range hopping in a general lattice model is of particular interest in the exploration of exotic quantum phases and QPTs. For example, as demonstrated in Appendix~\ref{appendix B}, our QRS model can be mapped to a frustrated $J_1$-$J_2$ spin model that has been intensively investigated in quantum magnetism. Despite a few decades of active research~\cite{PhysRevB.73.184420,PhysRevB.88.165138,PhysRevLett.128.227202}, key questions such as the existence of a spin liquid phase remain unanswered. Further studies of the QRS model, especially in one-dimensional and two-dimensional lattices, may shed new light on the investigation on the $J_1$-$J_2$ spin model.

	\begin{acknowledgments}
		This work is supported by the National Natural Science Foundation of China (Grants Nos. 12125402, 12074428, 92265208) and the National Key R\&D Program (Grant No. 2022YFA1405300). F.-X. S. acknowledges the China Postdoctoral Science Foundation (Grant No. 2020M680186). H.P. is supported by the US NSF PHY-2207283 and the Welch Foundation (Grant No. C-1669).  
	\end{acknowledgments}
	
		\appendix
		\section{Order parameters for SRP}\label{appendix A}
		 Notice that the first part of $H$ in Eq.~(\ref{QRM_SRP}) is similar to Eq.~(\ref{original Hamiltonian}), indicating that the Schrieffer-Wolff transformation can also be applied to project the Hamiltonian into the low energy spin subspace $P_-\equiv\ket{-}\bra{-}$. Typically, the local minimum can be determined by demanding  $V=0$. Considering the real and imaginary parts of $V$, the following equations can be derived from $\text{Re}(V)=\text{Im}(V)=0$
		\begin{align}
			&\omega A_n-\lambda \sin(2\gamma_n)+J_1(A_{n+1}+A_{n-1})+J_2A_{n+2}=0,
			\nonumber \\
			&\omega B_n+J_1(B_{n+1}+B_{n-1})+J_2B_{n+2}=0.
			\label{eq:imaginary-part}
		\end{align}
		Taking the periodic condition into consideration, we can easily get $B_1+B_3=B_2+B_4=0$, and $B_n=0$. Then the first line of Eq.~(\ref{eq:imaginary-part}) can be simplified as
		\begin{align}
			& \left(\omega-\dfrac{4\lambda^2}{\sqrt{16\lambda^2A_n^2+\Omega^2}}\right) A_n \nonumber \\
			& \hspace{1cm} +J_1(A_{n-1}+A_{n+1})+J_2A_{n+2}=0.
		\end{align}
		One solution is obtained by assuming $A_1=A_3=a$, $A_2=A_4=a'$, leading to
		\begin{align}
			\left(\omega+J_2-\dfrac{4\lambda^2}{\sqrt{16\lambda^2a^2+\Omega^2}}\right)a+2J_1a'&=0,
			\nonumber \\
			\left(\omega+J_2-\dfrac{4\lambda^2}{\sqrt{16\lambda^2a'^2+\Omega^2}}\right)a'+2J_1a&=0.
		\end{align}
		Then, we can obtain two solutions,
		\begin{align}
			a=a'=\pm\dfrac{1}{4\lambda}\sqrt{\dfrac{16\lambda^4}{(\omega+J_2+2J_1)^2}-\Omega^2},
			\nonumber \\
			a=-a'=\pm\dfrac{1}{4\lambda}\sqrt{\dfrac{16\lambda^4}{(\omega+J_2-2J_1)^2}-\Omega^2}.
			\label{eq:A1}
		\end{align}
		The critical point is reached when $a^2=a'^2=0$, giving $g_c=({1}/{2})\sqrt{1+{J_2}/{\omega}+{2J_1}/{\omega}}$ and $g_c=({1}/{2})\sqrt{1+{J_2}/{\omega}-{2J_1}/{\omega}}$, corresponding to the choice of $q=0$ an $\pi$ in Eq.~(\ref{QRM_critical_point}), respectively. The two-fold degenerate ground states in both cases indicate the breaking of $Z_2$ symmetry.
		
		Another case occurs when $A_1=A_4=a$, and $A_2=A_3=a'$. The conditions are as follows
		\begin{align}
			\left( \omega-\dfrac{4\lambda^2}{\sqrt{16\lambda^2a^2+\Omega^2}} \right) a+J_1(a+a')+J_2a'&=0,
			\nonumber \\
			\left( \omega-\dfrac{4\lambda^2}{\sqrt{16\lambda^2a'^2+\Omega^2}} \right) a'+J_1(a+a')+J_2a&=0.
		\end{align}
		If $a=a'$, the equations above will be reduced to the first case. The nontrivial solutions are
		\begin{align}
			a=-a'=\pm\dfrac{1}{4\lambda}\sqrt{\dfrac{16\lambda^4}{(\omega-J_2)^2}-\Omega^2}. \label{eq:A2}
		\end{align}
		Thus, the critical point can be obtained as $g_c=({1}/{2})\sqrt{1-J_2/\omega}$, which matches well with $g_c(q={\pi}/{2},{3\pi}/{2})$ in Eq.~(\ref{QRM_critical_point}). Notice that the ground state breaks both the $Z_2$ and $C_4$ symmetry simultaneously, which means the four-fold degenerate ground states occur if we consider the cyclic exchange of the four sites as $1234\rightarrow2341\rightarrow3412\rightarrow4123$. 
		
		\section{Validity analysis of the mean-field solution}\label{appendix}
		
		\begin{figure}[tb]
			\centering
			\subfigure{\includegraphics[width=0.45\textwidth]{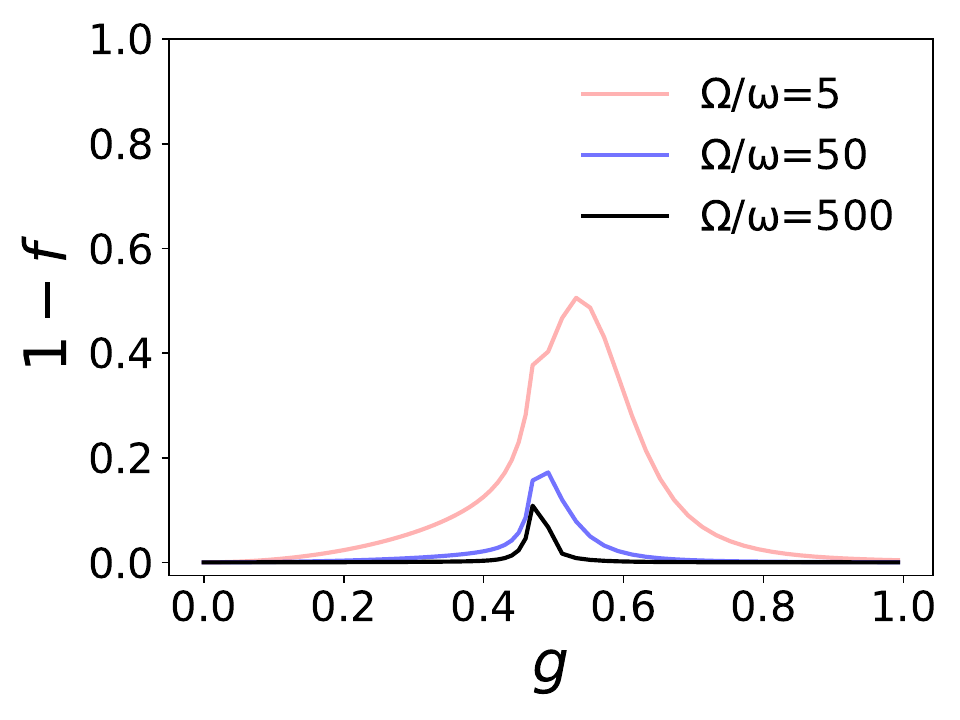}}
			\caption{The infidelity between the mean-field ground state $\psi_g$ and the numerical ground state $\psi_{\rm num}$ with fixed hopping rate $J_1=0.05$ and $J_2=0.07$ in our QRS model.}
			\label{error_numerical}
		\end{figure}

  The analytical results presented in the main text is based on the mean-field approximation. Here we will provide a quantitative comparison between the mean-field results and the fully numerical results based on the exact diagonalization of the original Hamiltonian (\ref{original Hamiltonian}). Specifically, we 
		define the fidelity for our mean-field ground state as $f\equiv\left|\langle\psi_g|\psi_{\rm num}\rangle\right|^2$, where $|\psi_{\rm num}\rangle$ is the ground state from the full numerical calculation, and 
		\begin{align}
			\ket{\psi_g}=D^\dagger(\vec{\alpha})\prod_{2q/{\pi}=0}^{3}S_q(\ket{0}\ket{-})^{\otimes4}
		\end{align}
		is the mean-field ground state.
		Notice that in the NP, $\vec{\alpha}=\vec{0}$, $\ket{-}=\ket{\downarrow}$, and the squeezed operator $S_q$ is defined in Sec. \ref{section}. While in the SRP, the squeezed parameter is changed as $\lambda_q[g]\rightarrow\lambda_q[g'={g_c^3(q)}/{g^2}]$.
		
		We illustrate in Fig.~\ref{error_numerical} the infidelity $1-f$ as a function of the coupling strength $g$. We find the infidelity is always small (i.e., $1-f\sim0$) far away from the critical point. By contrast, near the critical point, the infidelity increases sharply. However, this error can be greatly suppressed by increasing the ratio $\Omega/\omega$, suggesting that the mean-field becomes essentially exact in the thermodynamic limit $\Omega/\omega\rightarrow\infty$. We want to point out that the exact diagonalization near the critical point poses a challenging numerical problem, due to the divergent photon number. Here, we set a cut-off of the bosonic mode dimension as $N_c =5$ to guarantee convergence not too close to the critical point, which is sufficient to provide qualitative conclusions.
		
		\section{Mapping to spin $J_1$-$J_2$ model}\label{appendix B}
		By means of the Holstein-Primakoff transformation, given by $S_n^z=a_n^\dagger a_n-S_n$ and $S_n^+=a_n^\dagger\sqrt{2S_n-a_n^\dagger a_n}\approx a_n^\dagger\sqrt{2S_n}$, where the classic spin limit $S_n=S\rightarrow\infty$ is applied, the Hamiltonian Eq.~(\ref{Hamiltonian_SW}) can be mapped into a four-lattice $J_1$-$J_2$ model as 
		\begin{eqnarray}
			H_{J_1J_2}&=&\sum_{n=1}^4 \left[\omega S_n^z-\dfrac{2g^2\omega}{S}(S_n^x)^2\right]\notag\\
			&&+\dfrac{J_1}{S}\sum_{n=1}^4 \left(S_n^xS_{n+1}^x+S_n^yS_{n+1}^y\right)\notag\\
			&&+\dfrac{J_2}{2S}\sum_{n=1}^4 \left(S_n^xS_{n+2}^x+S_n^yS_{n+2}^y\right).
		\end{eqnarray}
		
		By defining $X_n\equiv\langle S_n^x\rangle/S$ and $Y_n\equiv\langle S_n^y\rangle/S$, the mean-field energy of such spin Hamiltonian is expressed as
		\begin{eqnarray}
			\dfrac{E_{J_1J_2}}{\omega S}&=&\sum_{n=1}^4 \Big[-\sqrt{1-X_n^2-Y_n^2}-2g^2X_n^2\notag\\
			&& \hspace{1cm} +J_1(X_nX_{n+1}+Y_nY_{n+1})\notag\\
			&& \hspace{1cm} +\dfrac{J_2}{2}(X_nX_{n+2}+Y_nY_{n+2})\Big].
		\end{eqnarray}
		A minimization of the energy $E_{J_1J_2}$ can give the similar branches of solutions for $\{ \langle\vec{S}_n\rangle \}$
		\begin{itemize}
			\item[(i)]\begin{align}
				X_n&=X_{n+1}=X_{n+2}
				\notag\\
				&=\pm\sqrt{1-\left(\dfrac{1}{4g^2-4[g_c(0)]^2+1}\right)^2},
			\end{align}
			\item[(ii)]\begin{align}
				X_n&=X_{n+2}=-X_{n+1}
				\notag\\
				&=\pm\sqrt{1-\left(\dfrac{1}{4g^2-4[g_c(\pi)]^2+1}\right)^2},
			\end{align} 
			\item[(iii)]\begin{align}
				X_n&=X_{n+1}=-X_{n+2}
				\notag\\
				&=\pm\sqrt{1-\left(\dfrac{1}{4g^2-4[g_c(\pi/2)]^2+1}\right)^2},
			\end{align} 
		\end{itemize}
		with the spin-$y$ components $Y_n=0$ in all the three cases. Obviously, the arrangements for spin-$x$ components $X_n$ correspond to the optical displacement $A_n$ in all the three different superradiant branches, shown as Eqs.~(\ref{eq:A1}) and (\ref{eq:A2}). And the dimensionless coupling strengths $g$ will be reduced to the critical points $g_c$ when $X_n=0$.
	
	\bibliography{ref}
	
\end{document}